\documentclass[%
 reprint,
 amsmath,amssymb,
 aps,
]{revtex4-2}

\usepackage{graphicx}% Include figure files
\usepackage{dcolumn}% Align table columns on decimal point
\usepackage{bm}% bold math
\usepackage[utf8]{inputenc} % allow utf-8 input
\usepackage[T1]{fontenc}    % use 8-bit T1 fonts
\usepackage{hyperref}       % hyperlinks
\usepackage{url}            % simple URL typesetting
\usepackage{booktabs}       % professional-quality tables
\usepackage{amsfonts}       % blackboard math symbols
\usepackage{nicefrac}       % compact symbols for 1/2, etc.
\usepackage{microtype}      % microtypography
\usepackage{xcolor}         % colors
\usepackage{sidecap}
\usepackage{multirow}
\usepackage[percent]{overpic}
\usepackage{float}
\usepackage{amsmath}
\usepackage{tikz}
\usetikzlibrary{bayesnet}

\usepackage{natbib}
\usepackage[export]{adjustbox}

\newcommand{\q}[1]{\lq #1\rq}
\newcommand{\trip}[1]{(x_#1, y_#1, z_#1)}

\newcommand{\x}{X}
\newcommand{\y}{Y}
\newcommand{\z}{Z}

\renewcommand{\v}{V}

\DeclareMathOperator{\vol}{vol}
\DeclareMathOperator{\VonMises}{VonMises}

\renewcommand{\citep}[1]{\cite{#1}}
\renewcommand{\citet}[1]{\cite{#1}}

\begin{document}

\preprint{APS/123-QED}

\title{Nearest-Neighbours Estimators for Conditional Mutual Information}% Force line breaks with \\

\author{Jake Witter}
% \email{first.Author@institution.edu}
\author{Conor Houghton}%
\affiliation{%
 School of Engineering Mathematics and Technology \\
  University of Bristol, Bristol, United Kingdom\\
  \texttt{(jake.witter / conor.houghton)@bristol.ac.uk}
}%

% \collaboration{MUSO Collaboration}%\noaffiliation

%  \homepage{http://www.Second.institution.edu/~Charlie.Author}
% \affiliation{
%  Second institution and/or address\\
%  This line break forced% with \\
% }%
% \affiliation{
%  Third institution, the second for Charlie Author
% \affiliation{%
%  Authors' institution and/or address\\
%  This line break forced with \textbackslash\textbackslash
% }%

% \collaboration{CLEO Collaboration}%\noaffiliation

\date{\today}% It is always \today, today,

\begin{abstract}
The conditional mutual information quantifies the conditional dependence of two random variables. It has numerous applications; it forms, for example, part of the definition of transfer entropy, a common measure of the causal relationship between time series. It does, however, require a lot of data to estimate accurately and suffers the curse of dimensionality, limiting its application in machine learning and data science. However, the Kozachenko-Leonenko approach can address this problem: it is possible in this approach to define a nearest-neighbour estimator which depends only on the distance between data points and not on the dimension of the data. Furthermore, the bias can be calculated analytically for this estimator. Here this estimator is described and is tested on simulated data.
% \begin{description}
% Secondary publications and information retrieval purposes.
% You may use the \texttt{description} environment to structure your abstract;
% use the optional argument of the \verb+\item+ command to give the category of each item. 
% \end{description}
\end{abstract}

\maketitle

%\tableofcontents

\section{Introduction}

The mutual information is a fundamental and common approach to assessing the relationship between random variables. It calculates the amount of information the value of one random variable contains, on average, about the value of another. The conditional mutual information is a natural extension of this, it is the average information in one variable about another, given knowledge of a third.

The conditional mutual information has many applications. In particular it is used to calculate the transfer entropy \citep{Schreiber2000,Palus2001}; the transfer entropy is a powerful approach to measuring causality which extends in a model-free manner the insight provided by the autoregressive formulation of commonly-used Granger causality \citep{BarnettBarrettSeth2009}. Conditional mutual information is also used to calculate the interaction information, the natural generalization of mutual information to more than two random variables \citep{McGill1954, Bell2003}.

Mutual information is often difficult to calculate in a model-free way: there is typically a huge data requirement when estimating it and it suffers from the curse of dimensionality, with the data required increasing rapidly as the dimension increases. This often renders conditional mutual information and transfer entropy impractical for problems involving the sort of multidimensional variables typically encountered in scientific applications and in data science and machine learning. In \citet{TobinHoughton2013,Houghton2015} a Kozachenko-Leonenko (KL) \citep{KozachenkoLeonenko1987,KraskovEtAl2004} estimator is presented, this gives a metric-based method for estimating mutual information. Since it uses only the distance between data points it wards off the curse of dimensionality. Here, this metric-space KL approach is extended to estimating conditional mutual information. 

This gives a formula for mutual information similar in spirit to the one derived in \citet{KraskovEtAl2004}. They are both nearest neighbour formulae derived using the KL estimator. The estimator described in \citet{KraskovEtAl2004} has been adapted to conditional mutual information \citep{Frenzel2007,Lizier2014} and transfer entropy \citep{Gomez2015};  here these are referred to as the `KSG' estimators. The novel estimator that is introduced here is referred to as the `new' estimator.

There are two significant differences between the KSG and new estimators; the first is that the derivation of the new estimator relies only on metric space properties of the data. Secondly, the new estimator allows the bias to be calculated and the bias-correction gives a useful approach to finding a value for $h$, the number of nearest neighbours. Although we believe these differences are both advantages that the new estimator has over the KSG estimator, one of the aims in this paper is to use simulated data to compare different approaches to estimating conditional mutual information `in action'.

\section{Methods}
The conditional mutual information between variables $X$ and $Y$, conditioned on $Z$, is given by
% \begin{equation}
%     I(X,Y|Z) =
%     \int_{\mathcal{X}}dx\int_{\mathcal{Y}}dy\int_{\mathcal{Z}}dz \log \left[ \frac{p_{XY|Z}(x,y|z) }{p_{X|Z}(x|z) p_{Y|Z}(y|z)} \right] p_{XYZ}(x,y,z)
%     \label{eq:conditional-mutual-information}
% \end{equation}
%
\begin{widetext}
\begin{equation}
    I(X,Y|Z) = \int_{\mathcal{X}}dx\int_{\mathcal{Y}}dy\int_{\mathcal{Z}}dz \log \left[ \frac{p_{XY|Z}(x,y|z) }{p_{X|Z}(x|z) p_{Y|Z}(y|z)} \right] p_{XYZ}(x,y,z)
    \label{eq:conditional-mutual-information}
\end{equation}
\end{widetext}
where $X$, $Y$ and $Z$ take values in $\mathcal{X}$, $\mathcal{Y}$ and $\mathcal{Z}$. Writing the integral as an expectation value and expanding the probability mass function to give an equivalent form that proves more convenient:
\begin{equation}
    I(X,Y|Z) = \left \langle \log  \frac{p_{XYZ}(x,y,z) p_Z(z)}{p_{XZ}(x,z) p_{YZ}(y,z)} \right \rangle_{p_{XYZ}(x,y,z)}
    \label{eq:conditional-mutual-information-expectation}
\end{equation}
The challenge with calculating this quantity lies in estimating the probabilities. The  histogram method estimates these probability mass functions by constructing histograms and counting the data points in each bin to estimate the likelihood of the corresponding outcomes.  This will give the correct answer in the infinite limit of available data and infinitely small bin sizes, but with three random variables the number of bins needed to produce an accurate answer is often far larger than what the quantity of data will support. Of course, this problem becomes more acute if $X$, $Y$ or $Z$ is a vector; the required number of points increases as a power of the number of dimensions. In transfer entropy, the random variables $X$ and $Z$ are vectors over recent time steps, making high-dimensionality unavoidable.

\subsection{Estimator}

Here the KL approach is described. It is intended here to compare both the KSG and new estimators to a ground truth. By using simulated data, it is possible to generate enough data to accurately calculate conditional information using histogram-based methods. The new estimator is described in detail but there is only a short outline of the KSG estimator since the details can be found in \citet{KraskovEtAl2004,Frenzel2007,Lizier2014,Gomez2015}.

A KL approach exploits the proximity structure of the data. Typically this proximity structure is based on coordinate distances but for the new estimator it is only assumed that the random variables, $X$, $Y$ and $Z$ take values in metric spaces $\mathcal{X}$, $\mathcal{Y}$ and $\mathcal{Z}$. Let
\begin{equation}
  \mathcal{D} = \{ \trip{1}, \trip{2},\ ...\ , \trip{n} \}
\end{equation}
denote the data. Depending on the problem being modelled the outcomes $x_i$, $y_i$ or $z_i$ could be high-dimensional vectors, or even values in a space without a manifold structure. However it is assumed that it is possible to calculate distances on the three spaces. Furthermore each set of triples $\trip{i}$ is assumed to be drawn from $p_{XYZ}(x,y,z)$.
Following from the definition of the conditional mutual information, Eq.~\ref{eq:conditional-mutual-information-expectation}, $I(X,Y|Z)$ can be approximated, for suitably large $n$, by the Monte-Carlo estimate
  \begin{equation}
    I(X,Y|Z) \approx \frac{1}{n}\sum_{i=1}^n \log \left[ \frac{p_{XYZ}(x_i,y_i,z_i) p_Z(z_i)}{p_{XZ}(x_i,z_i) p_{YZ}(y_i,z_i)}\right].
    \label{eq:conditional-mutual-information-sum}
\end{equation}
This reduces the problem to estimating $p_{XYZ}(x_i,y_i,z_i)$, $p_{XZ}(x_i,z_i)$, $p_{YZ}(y_i,z_i)$ and $p_Z(z_i)$ at the data points in $\mathcal{D}$. 

In the KL approach the probability at a point is approximated by counting the number of points in a nearby region. For simplicity consider temporarily a single random variable $V$ with data $\{v_1,v_2,\ldots,v_n\}$. The KL estimate of $p_V(v_i)$ is
\begin{equation}
  p_V(v_i)\approx \frac{h_\v(v_i)}{n\vol{[B_\v(v_i)]}}
\end{equation}
where $B_V(v_i)$ is a region in $\mathcal{V}$ around $v_i$, $h_\v(v_i)=|B_\v(v_i)|$
is the number of data points in the region and $\vol{[B_\v(v_i)]}$ is the volume of the region $B_\v(v_i)$. The estimate follows from the definition of the probability mass $p_V(v)$:
\begin{equation}
  \langle h_\v(v_i)\rangle=n\int_{B_\v(v_i)}dv\, p_V(v) 
\end{equation}
by using two approximations. Firstly approximating $p_V(v)$ by assuming it is constant throughout $B_\v(v_i)$ and secondly, approximately evaluating the integral by counting the data points: $h_\v(v_i)\approx\langle h_\v(v_i)\rangle$. Clearly these two approximations favour different sizes of $B_\v(v_i)$: approximating $p_V(v)$ as constant favours a small region, approximating the expected number of data points by the actual size of the sample found in $B_\v(v_i)$ favours a large region.

The other important ingredient is the calculation of the volume $\vol{B_\v(v_i)}$. Since this is a metric space approach, to avoid any assumptions about a coordinate structure, this calculation follows \cite{Houghton2015} and calculates volumes using probability densities.  In essence, the volume of a region is given by the probability mass it contains. This is easier to discuss in the context where there are different probability densities, as in the problem being considered here where the goal is to estimate conditional information and this requires the estimate of a number of different probability mass functions: $p_{XYZ}(x_i,y_i,z_i)$, $p_{XZ}(x_i,z_i)$, $p_{YZ}(y_i,z_i)$ and $p_Z(z_i)$. In fact, the key to estimating the volume is to use different distributions on the same space. This is required because using the same distribution to estimate volume and probability mass trivializes the estimate: to see this consider, for example, $p_Z(z_i)$. To define the region $B_\z(z_i)$ a volume is fixed, say $h/n$; where $h$ counts the number of points and so can be thought of as a smoothing parameter. Now, for each point $z_i$ a ball is expanded around this seed point until it has volume $h/n$, since the probability mass is being used to measure the volume, this is estimated by counting data points:
\begin{equation}
  \vol{B_z(i)}=\frac{1}{n}\int_{B_{\z(i)}}dz\,p_Z(z) =\frac{\langle h_\z(i)\rangle}{n}\approx \frac{h_\z(i)}{n}
\end{equation}
where, for brevity, the shorthand $B_\z(i)=B_\z(z_i)$ and
$h_\z(i)=h_\z(z_i)$ has been used.  Thus, using the metric, the region $B_z{i}$ is a ball in
$\mathcal{Z}$ which has been expanded around the seed point until it contains $h$ points, including $z_i$ itself. Now, using the KL estimate of the probability:
\begin{equation}
  p_Z(z_i)\approx \frac{h_\z(i)}{n\vol{B_\z(i)}}=1,
\end{equation}
and thus gives a trivial estimate.

\begin{figure*}[t]
  \begin{center}
    \includegraphics{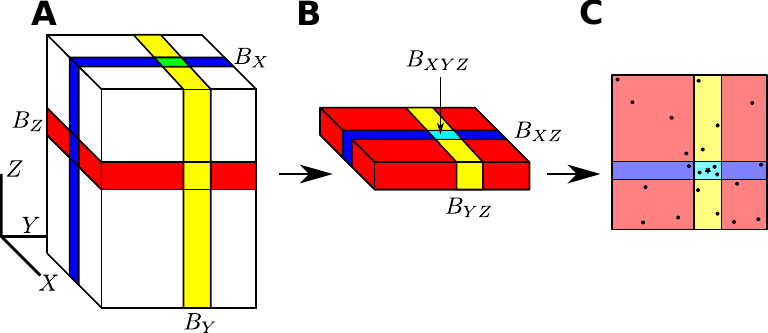}
  \end{center}
  \caption{\textbf{An illustration of the different intersections}. In
    this cartoon the three sets of outcomes, $\mathcal{X}$,
    $\mathcal{Y}$ and $\mathcal{Z}$, are marked $X$, $Y$ and $Z$; here these are one-dimensional whereas, of course, they can be high-dimensional or even spaces without a dimension. In \textbf{A} the three balls around the seed point, $B_{\x}$, $B_{\y}$ and $B_{\z}$, are drawn in blue, yellow and red. Each of these balls contain $h$ points, this determines their radius. In \textbf{B} attention is restricted $B_{\z}$; the intersection this has with $B_{\x}$ gives $B_{\x\z}$ and with $B_{\y}$ gives $B_{\y\z}$. All three intersect to give the cyan region $B_{\x\y\z}$. \textbf{C} shows $B_{\z}$ again, from above; the points are also marked, with a star for the seed point. Here, $h_{\x\y\z}=4$, $h_{\x\z}=6$, $h_{\y\z}=9$ and all the points lie in $B_{\z}$ so $h$ in this illustration is 22.\label{fig:squares}}
\end{figure*}

To avoid this problem a different distribution is used to estimate volumes. For joint distributions this is provided by the marginalized distribution. First, consider $p_{XZ}(x_i,z_i)$; in this case $p_X(x_i)p_Z(z_i)$ can be used to measure the volume. At first this is confusing, the volume of the ball around $(x_i,z_i)$ depends on the number of $x_j$ points and the number of $z_j$ points, not on the number of $(x_j,z_j)$ pairs, as it would if $p_{XZ}(x,z)$ was being used as the volume measure. As such, the ball in $\mathcal{X}\times\mathcal{Z}$ is formed by expanding a ball around $x_i$ until it contains $h$ points and a ball around $z_i$, again, until it also contains $h$ points. The ball $B_{\x\z}(x_i,z_i)$ is then the intersection of these two balls; it has volume estimated as $h^2/n^2$. Writing $h_{\x\z}(i)$ as the number of points it contains, the KL estimate is
\begin{equation}
  p_{XZ}(x_i,z_i)\approx \frac{nh_{\x\z}(x_i,z_i)}{h^2}.
\end{equation}

In the case of $p_{XYZ}(x,y,z)$ the alternative distribution on the same space is again provided by the marginal distribution, $p_X(x)p_Y(y)p_Z(z)$ and the ball $B_{\x\y\z}(x_i,y_i,z_i)$ is formed the same way as for the $p_{XZ}(x,z)$ example. It is the intersection of $B_\x(i)$, $B_\y(i)$ and $B_\z(i)$ and has volume $h^3/n^3$. This gives
\begin{equation}
  p_{XYZ}(x_i,y_i,z_i)\approx \frac{n^2h_{\x\y\z}(i)}{h^3}
\end{equation}
where $h_{\x\y\z}(i)$ is the number of points in $B_{\x\y\z}(x_i,y_i,z_i)$. The regions have been referred to as balls, though, in some ways, they are more like squares or cubes; Fig.~\ref{fig:squares} provides an illustration. From now on, the shorthand introduced earlier is extended to these regions, so $B_{\x\y\z}(i)=B_{\x\y\z}\trip{i}$ and so on.

Using these point probability estimates, the conditional mutual information, Eq.~\ref{eq:conditional-mutual-information-sum}, can be rewritten as
\begin{equation}
    \begin{split}
    I(X,Y|Z) \approx\frac{1}{n}\sum_{i=1}^n\log \frac{h_{\x\y\z}(i) h}{h_{\x\z}(i)h_{\y\z}(i)}. 
    \end{split}
    \label{eq:rewritten-once}
\end{equation}
Thus, the conditional mutual information is estimated by finding the nearest-neighbours for each data point in each of $\mathcal{X}$, $\mathcal{Y}$ and $\mathcal{Z}$ and then counting the points in the intersections $B_{\x\z}(i)$, $B_{\y\z}(i)$ and $B_{\x\y\z}(i)$. One complication arises when there are draws; this is discussed in Appendix 1.

\subsection{Bias}

As is typical with estimators of mutual information, this approach has an upward bias. That is, when applying the approach to conditionally independent $X$ and $Y$:
\begin{equation}
  X\rightarrow Z \rightarrow Y
\end{equation}
there will be a non-zero estimated conditional mutual information, when, of course, the true value is zero. To remove this bias the expected value of the estimator is calculated, assuming $X$ and $Y$ are conditionally independent, given $Z$. This follows similar calculations in \cite{Houghton2019,WitterHoughton2021} but is more complicated in this case because of the additional random variable.

Calculating the bias requires a formula for the probability that the intersection $B_{\x\y\z}(i)$ contains $r$ points given the values $h_{\x\z}(i)$ and  $h_{\y\z}(i)$, that is the sizes of $B_{\x\z}(i)$ and $B_{\y\z}(i)$, something that will vary from point to point. This bias therefore varies for each data point triple, each $i$, unlike similar formulae for the bias of the estimator in \cite{Houghton2019,WitterHoughton2021}. The reason for this difference is that the independence is conditional and so the bias has to include the dependence that $X$ and $Y$ might have while calculating the expected size of $B_{\x\y\z}(i)$ under the assumption of conditional independence. Thus, the bias includes a sum over all data points, for $(x_i,y_i,z_i)$ the contribution to the bias is
\begin{equation}
   I_b(i,h) = \sum^h_{r=1} \mathbb{P}(h_{\x\y\z}(i) = r) \log \frac{r h}{h_{\x\z}(i) h_{\y\z}(i)}.
\end{equation}

In order to calculate $\mathbb{P}(h_{\x\y\z}(i) = r)$ consider the $h_{\x\z}(i)$ points in $B_{\x\z}(i)$; these points are chosen from the $h$ points in $B_\z(i)$. The number of these points that are also among the $h_{\y\z}(i)$ points in $B_{\y\z}(i)$ gives $r$. This is an urn problem. First remove the seed point $\trip{i}$ itself, now there are $h-1$ balls in an urn, $h_{\x\z}(i)-1$ of which are \q{red}, meaning they are in $B_{\x\z}(i)$. Next $h_{\y\z}(i)-1$ balls are drawn, representing the number in $B_{\y\z}(i)$, $r-1$ is the number of these balls that are
\q{red}. This is described by the hypergeometric distribution, so the probability $p(r)$ that $h_{\x\y\z}(i) = r$ is
\begin{align}
        p(r) &= \frac{\left(\begin{array}{c} h_{\x\z}(i)-1 \\ r-1 \end{array}\right)\left(\begin{array}{c} h-h_{\x\z}(i) \\ h_{\y\z}(i)-r \end{array}\right)}{\left(\begin{array}{c} h-1 \\ h_{\y\z}(i)-1 \end{array}\right)}  \\  
    &\equiv \mbox{Hypergeometric}(h-1, h_{\x\z}(i)-1, h_{\y\z}(i)-1).\nonumber
\end{align}
Now, it is possible to calculate $I_b(i,h)$ and hence the total bias
\begin{equation}
  I_b(h)=\frac{1}{n}\sum_{i=1}^n I_b(i,h)
\end{equation}
It is clear that the result of the estimator will be a function of $h$. As mentioned above, a larger $h$ will mean a larger region, increasing the error introduced by assuming the probability density is constant throughout the region, but, conversely, makes it more likely that the regions contain a representative number of points. While it is unclear how to choose $h$, practically, maximising the bias-corrected information over $h$ works well and gives a good estimate of the information, as in \cite{Houghton2019}; it balances the two sources of error, both of which might be expected to reduce the de-biased estimate. Hence
%\begin{widetext}
\begin{equation}
    I(X,Y|Z) \approx
    \max_h \frac{1}{n} 
    \sum_{i=1}^{n} \log \frac{h_{\x\y\z}(i) h}{h_{\x\z}(i)h_{\y\z}(i)} 
     - I_b(h)
    \label{eq:full}
\end{equation}
%\end{widetext}
Thus, the procedure is to maximize the bias-corrected mutual information over $h$; this is performed here using golden section search \cite{PressEtAl2007}. This procedure does cause an upward bias when the mutual information is low: when there is little mutual information the noise in the estimate is larger compared to the true value and so maximizing over $h$ effectively selects the value of $h$ whose corresponding estimate is largest because of the noise, rather than because this is the optimal value of $h$. However, for the more typical case, where the estimator is being used with variables that are not conditionally independent, the procedure appears to work well.

\subsection{The KSG estimator}

%check Wibral et al citation from the Lizier ppaer, near B6

%what is below doesn't tell the full story, there needs to be some comparison between the X and Y %spaces!
%this was always the complication with "outer" counts, it is why we abandoned it in our case!

The KSG estimator, \citep{KraskovEtAl2004} also adopts a KL approach in that it also estimates the probabilities by counting data points in small regions. The most obvious difference between the KSG estimator and the new estimator lies in the calculation of volumes. In the new estimator, as described above, the volume is also estimated by counting data points whereas the KSG estimator assumes a coordinate system and calculates volumes by integration. There is a second difference, in the KSG estimator the probability density is estimated by calculating the probability of finding the $k$th point at a particular distance, a more sophisticated approach to the one used in the new estimator, but one that depends on the space having a coordinate system relevant to the calculation of mutual information. Ultimately this leads to a formula that contains a digamma function, \cite{AbramowitzStegun1968}, rather than a logarithm, the digamma function is typically denoted  $\psi(x)$. 

There are two, similar, versions of the KSG estimator:
\begin{align}\label{KSG1}
I(X,Y)\approx&\psi(k)+\psi(n)\\&-\frac{1}{n}\sum_{i}[\psi(k_X(i)+1)+\psi(k_Y(i)+1)] \nonumber   
\end{align}
and
\begin{align}\label{KSG2}
I(X,Y)\approx&\psi(k)-\frac{1}{k}+\psi(n)\\&-\frac{1}{n}\sum_{i}[\psi(k_X(i))+\psi(k_Y(i))]    \nonumber
\end{align}
where $k$ is a smoothing parameter, similar to $h$ in the new estimator and $n$ is the total number of points. Like the new estimator, the formulae involve counting points, with  $k_X(i)$ and $k_Y(i)$ giving the number of points in particular regions. They are, however, potentially confusing in that there is actually a subtly different definition used for each of the two formula. The definition is also different from the one used for quantities such as $h_{XY}(i)$ in the new estimator. $k_X(i)$ and $k_Y(i)$ are what could be thought of as ``outer'' counts whereas $h_{XY}(i)$ is an ``inner'' count. Loosely, in an inner count two regions, $B_X(i)$ and $B_Y(i)$, are given with specified size $h$ and the number of points in the intersection, $h_{XY}(i)$, is calculated. In an outer count this is reversed: the size of an intersection, $k$, is specified and counting is performed on the two corresponding regions. More specifically, to form the outer count used in the KSG estimator, the distance in the product space $\cal{X}\times\cal{Y}$ from a seed point $(x_i,y_i)$ to its $k$th nearest neighbour is found. This defines the intersection. The subtly arises because this requires a metric on $\cal{X}\times\cal{Y}$ when the metrics are only defined on $\cal{X}$ and $\cal{Y}$ separately. To lift these metrics to a metric on the product, the maximum is used, with the obvious notation
\begin{equation}
    d_{XY}[(x_i,y_i),(x_j,y_j)]=\text{max}[d_X(x_i,x_j),d_Y(y_i,y_j)]
\end{equation}
Let $d^*$ denote the distance to the $k$th nearest point. There are now two options, the first uses this distance to define $B_X(i)$ and $B_Y(i)$, so, for example, $B_X(i)$ is the set of all points $x_j$ within $d^*$ of the seed point $x_i$, that is $B_X(i)=\{x_j:d_X(x_i,x_j)<d^*\}$. The second does something more complicated, it uses $d^*_X$ to define $B_X(i)$ and $d^*_Y$ to define $B_Y(i)$ where $d^*=\text{max}[d_X^*,d_Y^*]$. In either case, the outer counts are then $k_X(i)=|B_X(i)|$ and $k_Y(i)=|B_Y(i)|$. The two KSG estimators above, Eq.~\ref{KSG1} and Eq.~\ref{KSG2}, correspond to these two options.

It is interesting that these formula do not include terms that depend on the dimension and in this way, although the derivation of the KSG estimator relies on the existence of coordinates, the formula itself does not. Beyond the theoretical objection that these estimators are not derived for spaces without coordinates, even if they can be applied to spaces of that type, the use of the maximum metric on $\mathcal{X}\times\mathcal{Y}$ creates practical difficulties in using these estimators relative to using the new estimator. Most obviously it relies on some compatibility of the metrics on $\mathcal{X}$ or $\mathcal{Y}$ since they are compared when calculating the distance on $\mathcal{X}\times\mathcal{Y}$. There is a milder version of this issue in the case of the new estimator where the choice of the same volume for both $B_X(i)$ and $B_Y(i)$ also makes an implicit assumption of similarity. The other challenge is that it is more difficult to calculate the bias.  A consequence of using the maximum metric is that counts $k_X(i)$ and $k_Y(i)$ depend on the distributions of points in $\cal{X}$ and $\cal{Y}$ even if these two distributions are independent, preventing the calculation of the bias without making additional assumptions.

%KSG is not biased; work out why?

These estimators are in turn used to derive estimators of the conditional mutual information in \cite{Frenzel2007,Lizier2014} and the transfer entropy in \cite{Gomez2015}. These estimators, like the new estimator proposed here, are similar to the mutual information estimators but contain more complicated counts of data points. The two estimators are
\begin{widetext}
\begin{equation}
I(X,Y|Z)\approx\psi(k)+\frac{1}{n}\sum_{i}[\psi(k_Z+1)-\psi(k_{XZ}(i)+1)-\psi(k_{YZ}(i)+1)]    
\end{equation}
\begin{align}
I(X,Y|Z)\approx&\psi(k)-\frac{2}{k}+\frac{1}{n}\sum_{i}\left[\psi(k_Z(i))-\psi(k_{XZ}(i))-\frac{1}{k_{XZ}(i)}-\psi(k_{YZ}(i))-\frac{1}{k_{YZ}(i)}\right]    
\end{align}
\end{widetext}
where, again, variables like $k_{XZ}$ represents an outer count. In short the $B_X(i)$, $B_Y(i)$ and $B_Z(i)$ balls are expanded to include $k$ points, $k_Z(i)=|B_Z(i)|$ whereas $k_{XZ}(i)$ is the number of $(x,z)$ data points in $B_X(i)\times B_Z(i)$.

%These estimators differ from the new estimator in relying on coordinates in their derivation, if not their implementation; the derivation is, in general, more subtle and the estimation of the bias for zero information is not provided, nor is there an approach to selecting a value of $K$.

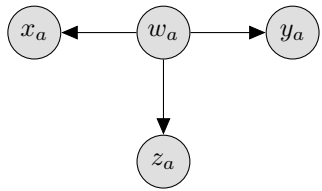
\begin{figure}[tb]
    \centering
    \begin{tikzpicture}
        \node[obs]                            (z) {$z_a$};
        \node[obs, above=of z]                (w) {$w_a$};   
        \node[obs, left=of w]                 (x) {$x_a$};
        \node[obs, right=of w]                (y) {$y_a$};
        \edge {w} {x};
        \edge {w} {y};
        \edge {w} {z};
    \end{tikzpicture}
    \caption{\textbf{The Markov network for our simulated data}. This illustrates the relationship between the three observed variables $X$, $Y$ and $Z$ and the unobserved variable $W$. $X$ and $Y$ are conditionally independent given $W$; since $Z$ is a noisy version of $W$ the mutual independence of $X$ and $Y$ given $Z$ depends on the amount of that noise, $\sigma_z$.  \label{fig:markov}}
\end{figure}

\subsection{Transfer Entropy}

One important application of this new estimator for conditional mutual information is the calculation of transfer entropy: a directed measure of information transfer between two processes \cite{Palus2001}. This quantity can be useful as it gives a direction to the relationship between processes, as well as an insight in to whether a relationship is causal. Consider $X_t$, $Y_t$ to be two random processes with $t \in \mathbb{N}$ a discrete time index. The transfer entropy from process $X$ to process $Y$ is the mutual information between past values of $X$ and the present value of $Y$, conditioned on the past values of $Y$. That is, as described in \cite{HlavavckovaEtAl2007},
\begin{equation}
  T(X \rightarrow Y) = I(Y_t , X_{t-1:t-\ell} | Y_{t-1:t-\ell})
  \label{eq:TE}
\end{equation}
where $X_{t-1:t-\ell}=(X_{t-1},X_{t-2},\ldots,X_{t-\ell})$ is the past $\ell$ values of $X_t$, with a similar notation for $Y_{t-1:t-\ell}$. Thus, roughly speaking, the transfer entropy measures the extra information about the present value of $Y$ contained in the past values of $X$ beyond what is already accounted for in the past of $Y$. Transfer entropy is widely used in neuroscience, see for example \cite{VicenteEtAl2011,StaniekLehnertz2008} and finance, see \cite{BossomaierEtAl2016}, but the data requirement for estimating conditional mutual information is an impediment to its practical use. The transfer entropy can be thought of as a non-parametric version of the commonly-used Granger causality; or, rather, transfer entropy can be considered the ideal approach to estimating causal relationship when there is no data-constraint, with the Granger causality interpreted as an approximation to the transfer entropy which is exact for Gaussian random variables \cite{BarnettBarrettSeth2009}.

\section{Example}

\subsection{A simple Markov tree}

To demonstrate the estimation of conditional mutual information and to compare approaches, a simple example is considered here. This example is designed to produce data with a conditional mutual information that can be varied by changing a parameter. It is also designed to allow for the easy generation of a large quantity of data, which means that histogram methods can be used to give something approaching a ground truth. The example consists of four Gaussian random variables; $W$, $X$, $Y$ and $Z$, distributed as
\begin{align}
    w_a & \sim \text{Normal}(0, \sigma_w) \cr
    x_a & \sim \text{Normal}(w_a, \sigma_x) \cr
    y_a & \sim \text{Normal}(w_a, \sigma_y) \cr
    z_a & \sim \text{Normal}(w_a, \sigma_z)
\end{align}
where the $a$-index corresponds to the different dimensions of the data, so, for example, for two dimensions $X=(X_1,X_2)$. Here one- and two-dimensional examples will be considered. This relationship is illustrated in Fig.~\ref{fig:markov}.

\begin{figure*}
    \centering
    \begin{overpic}[width=0.7\textwidth,tics=10]{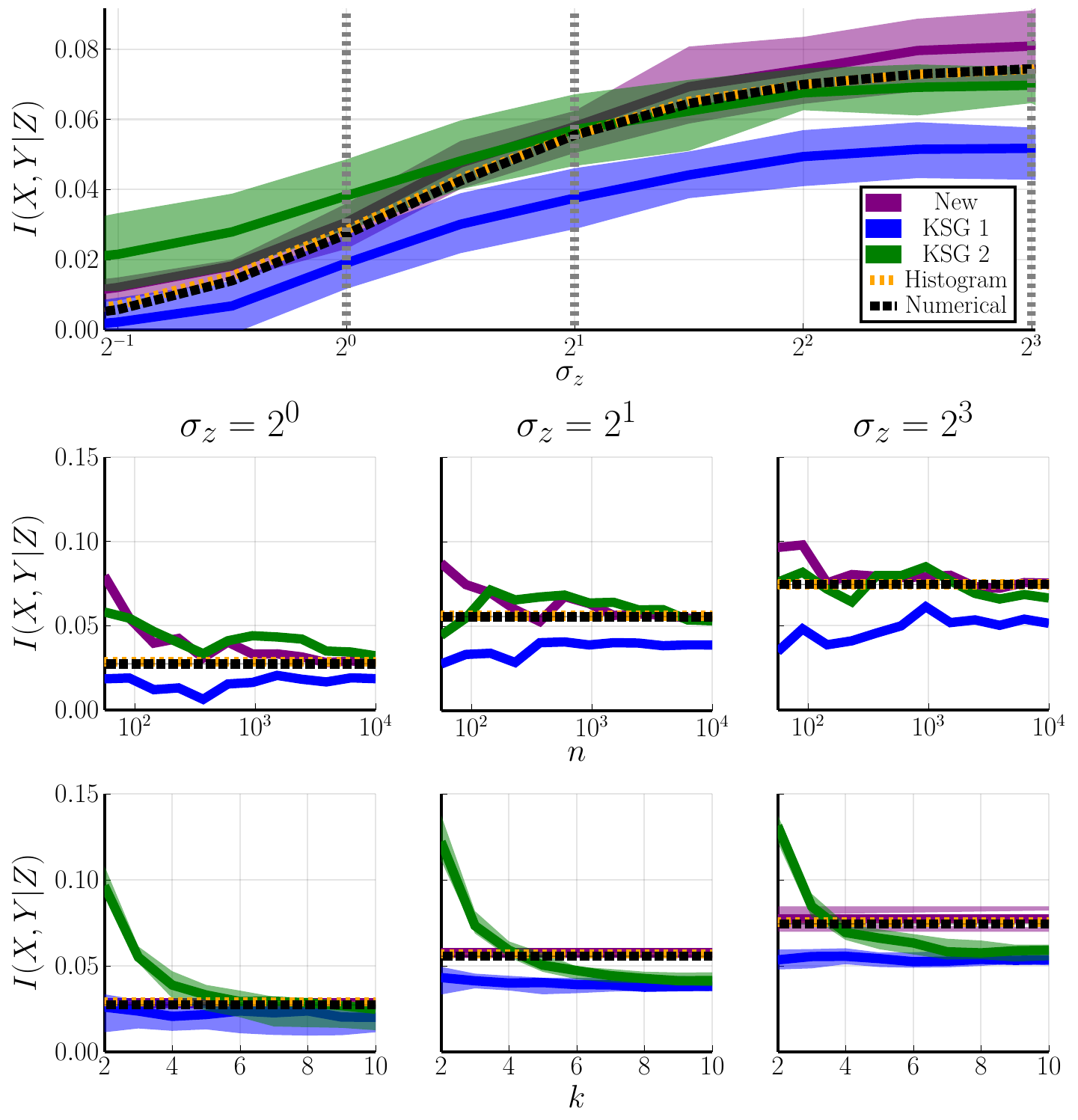}
        \put (0,98) {\huge \textbf{a}}
        \put (0,58) {\huge \textbf{b}}
        \put (0,28) {\huge \textbf{c}}
    \end{overpic}
    \caption{\textbf{Results from one-dimensional Markov tree.} In \textbf{a} the relationship between the estimated conditional mutual information and $\sigma_z$ is shown. The shaded area shows the middle 50\% of estimates. Here, the KSG and new methods use 3500 points, while the histogram uses $5 \times 10^6$ points to establish the ground truth. \textbf{b} shows how this estimate scales over the number of data points used in the  estimate. Again, the histogram method uses $5 \times 10^6$ points. In \textbf{c}, the relationship between the $k$ used in the KSG estimates, and the estimated information is shown. This demonstrates that choice of $k$ does strongly influence the estimated information. Note the change in scale in the vertical axis between \textbf{a} and the other two rows.}
    \label{fig:markov-tree-results-1d}
\end{figure*}

\begin{figure*}
    \centering
    \begin{overpic}[width=0.7\textwidth,tics=10]{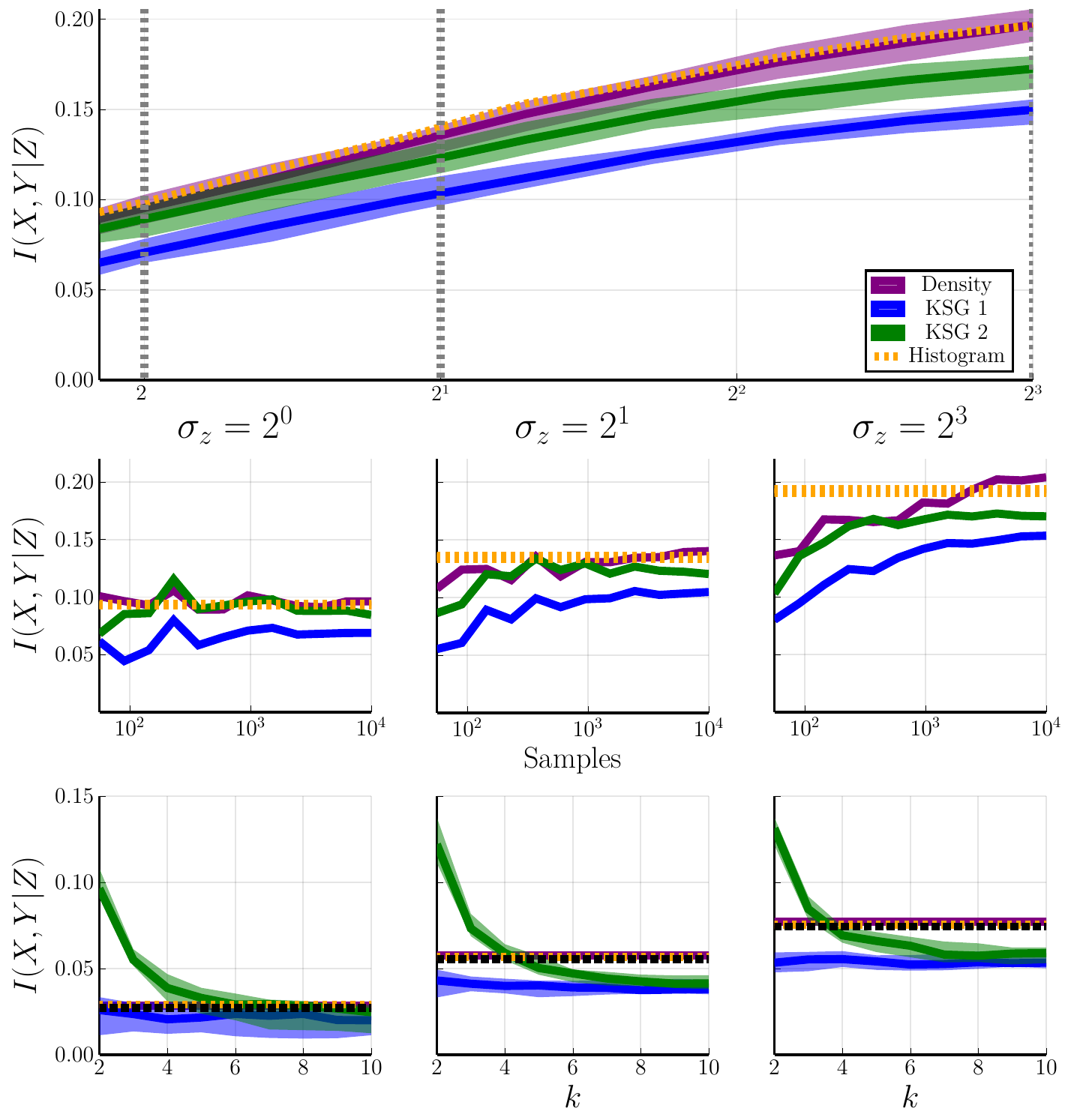}
        \put (0,98) {\huge \textbf{a}}
        \put (0,58) {\huge \textbf{b}}
        \put (0,28) {\huge \textbf{c}}
    \end{overpic}
    \caption{\textbf{Results from two dimensional Markov tree} In \textbf{a} the relationship between the estimated conditional mutual information and $\sigma_z$ is shown. The shaded area shows the middle 50\% of estimates. Here, the KSG and new methods use 3500 points, while the histogram uses $5 \times 10^6$ points to establish the ground truth. \textbf{b} shows how this estimate scales over the number of data points used in the  estimate. Again, the histogram method uses $5 \times 10^6$ points. In \textbf{c}, the relationship between the $k$ used in the KSG estimates, and the estimated information is shown. This demonstrates that choice of $k$ does strongly influence the estimated information. Note the change in scale in the vertical axis between \textbf{c} and the other two rows.}
    \label{fig:markov-tree-results-2d}
\end{figure*}

Clearly $X$ and $Y$ are related, $I(X, Y) > 0$, but they are conditionally independent when conditioned on $W$, so $I(X,Y|W)=0$. However, here $I(X,Y|Z)$ is estimated and $Z$ is not $W$, in fact $Z$ can be thought of as a noisy version of $W$. Thus, $I(X,Y|Z)$ is non-zero and will depend on how noisy $Z$ is, as determined by $\sigma_z$. As $\sigma_z$ approaches zero, $Z$ becomes closer to an exact copy of $W$ and $I(X,Y|Z)$ approaches zero. Conversely, as $\sigma_z$ becomes large, conditioning on $Z$ becomes increasingly meaningless and $I(X,Y|Z)$ approaches $I(X,Y)$. The estimator is applied to the example in one and two dimensions, as seen in Fig.~\ref{fig:markov-tree-results-1d} and Fig.~\ref{fig:markov-tree-results-2d}. 

Considering first the one-dimensional case, Fig.~\ref{fig:markov-tree-results-1d}\textbf{a} graphs the mutual information as $\sigma_z$ varies, the histogram estimate and a numerical show show the mutual information increasing, as expected, as $\sigma_z$ increased. For the estimators a fixed $n$ of 3500; while all three give a reasonable estimate, the new estimator is closer to the correct value. In Fig.~\ref{fig:markov-tree-results-1d}\textbf{b} the behaviour of the estimators for differing value of $n$ is considered for three different values of $\sigma_z$, this shows that all three estimators appear to require about 3500 samples to reach their best estimate. The two KSG estimators require a choice of smoothing parameter $k$, in Fig.~\ref{fig:markov-tree-results-1d}\textbf{c} it is seen that this choice does affect the estimate for KSG2 and the estimated value varies from too large to too small as $k$ increases; KSG1 consistently gives too small a value.

The same set of measures are exhibited for two dimension in Fig.~\ref{fig:markov-tree-results-2d}. Here, the results are somewhat similar, but more extreme. That is, the advantage of the new estimator becomes more pronounced, the advantage of using the new estimator becomes more extreme and convergence to the best estimate as the number of sample $n$ increases is slower in all cases.

The histogram approach was applied with a huge number of data points; this indicates that the histogram approach and the estimator described in this paper approach similar values, but vastly more data is required for the histogram approach. This discrepancy is more pronounced in the two-dimensional example, and, in fact, becomes even more so as dimension increases further. 

\subsection{XY Model}

In order to assess the ability of the new estimator to calculate transfer entropy, a model is needed which combines noisy dynamics and causal relationships. Here, the XY model is used.

%\begin{figure}[tbh]
%    \centering
%    \includegraphics[width=0.33\textwidth]{figs/frame.pdf}
%    \caption{\textbf{XY model example}. An example state of the XY model as described here. The fixed pixel is in the centre, at $(25, 25)$, with colour representing the value of $\theta$ at each site.}
%    \label{fig:xy-model-frame}
%\end{figure}

The XY model is a well studied model in statistical mechanics which gives a mathematical description of interacting spins. It is a lattice model in which individual particles at lattice points have a spin described by a point on a circle. The spins of adjacent particles interact in a way that models both the tendency of spins to align and thermodynamics effects which allow for fluctuations away from alignment. Here, a two-dimensional square-lattice XY model is considered, with periodic boundary conditions in both directions, giving a torus. Using
\begin{equation}
    \textbf{s}_x = (\cos\theta_x,\ \sin\theta_x)
\end{equation}
to denote the spin at the lattice point $x$, the energy is given
\begin{align}
    H(\sigma) &= - J \sum_{x\in\mathcal{G}}\sum_{y\in\mathcal{N}(x)} \mathbf{s}_x\cdot\mathbf{s}_y\nonumber\\&=- J \sum_{x\in\mathcal{G}}\sum_{y\in\mathcal{N}(x)}\cos (\theta_x - \theta_y)
\end{align}
where $\mathcal{N}(x)$ is the set made up of the four neighbours to $x$ and $\mathcal{G}$ is the set of all points in the lattice. This corresponds the simplest version of the model where coupling strengths are constant and there is no external field. Here the `Metropolis' version of the dynamics is used, this is one of a number of different, ultimately equivalent, approaches to the statistical dynamics. In this implementation, at each time step, all sites are potentially updated. A change is proposed for a site by drawing a new value from the von Mises distribution centered on the current value:
\begin{equation}
    \theta'_x \sim \VonMises(\theta_x, r)
\end{equation}
where $r$ is a concentration parameter; for small $r$ the distribution approaches a uniform distribution, for large $r$ the value of $\theta'_x$ will tend to be very close to the value of $\theta_x$. Now, the change in energy $\delta H$ that would result from switching $\theta_x$ to the proposed new value $\theta'_x$ is calculated. If $\delta H<0$ the proposal is accepted corresponding to the dynamical tendency of spins to align. If, however, $\delta H>0$ the proposal may still be accepted with probability $\exp{(-\delta H/T)}$; this corresponds to thermal fluctuations. $T$ is a temperature, an adjustable parameter which determines the significance of the thermal effects. In practice, $r$, the value of the concentration in the proposal distribution, is varied to keep the overall acceptance rate equal to some target value $d$. In simulations, this is taken to be $0.5$.

Here, the focus is on accessing causal structure. To this end, the XY-model has been modified so there is one site on the grid that affects the dynamics of the other sites, but is itself unaffected by its neighbours. In short, this special site, which will be referred to as the causal site, changes according to a Bernoulli random walk, it changes its angle by $\pm 0.2$ at each time step. This is independent of the states of the surrounding sites, however the causal point still \q{informs} its neighbour in that it is incorporated in the usual way in their update.

% \begin{figure}
%     \centering
%     % \includegraphics[width=0.28\textwidth]{example-image-a}
%     % \includegraphics[width=0.25\textwidth]{figs/XYModel_frame.pdf}
%     \includegraphics[width=0.3\textwidth]{figs/TE_over_distance.pdf}
%     \caption{
%     \textbf{Distance against transfer entropy}. An example illustrating the relationship between distance and calculated transfer entropy, in the XY model. Results are shown for varying past lengths.
%     }
%     \label{fig:xy-te-against-distance}
% \end{figure}

\begin{figure}
    \centering
    \begin{overpic}[width=0.23\textwidth,tics=10]{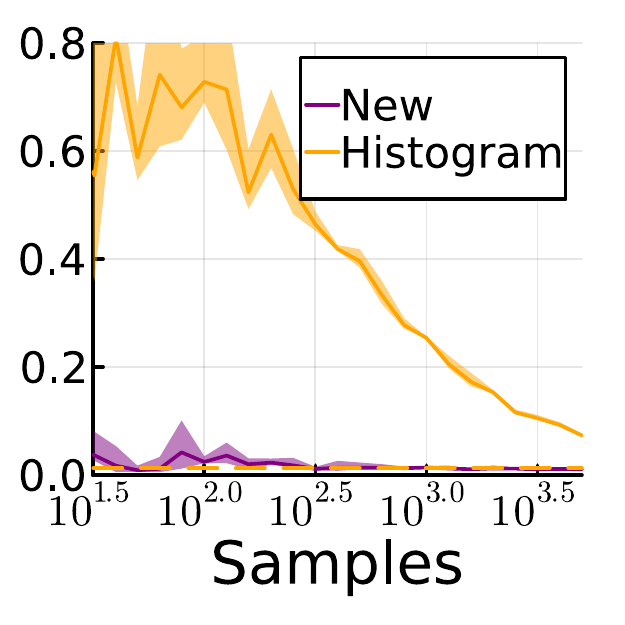}
        \put (-8,90) {\huge \textbf{a}}
    \end{overpic}
    \includegraphics[width=0.23\textwidth]{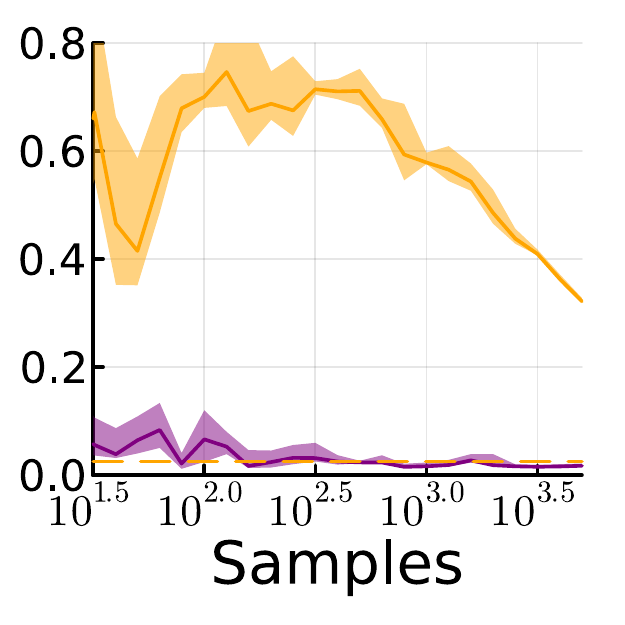}
    
    \vspace{0.4cm}
    
    \begin{overpic}[width=0.23\textwidth,tics=10]{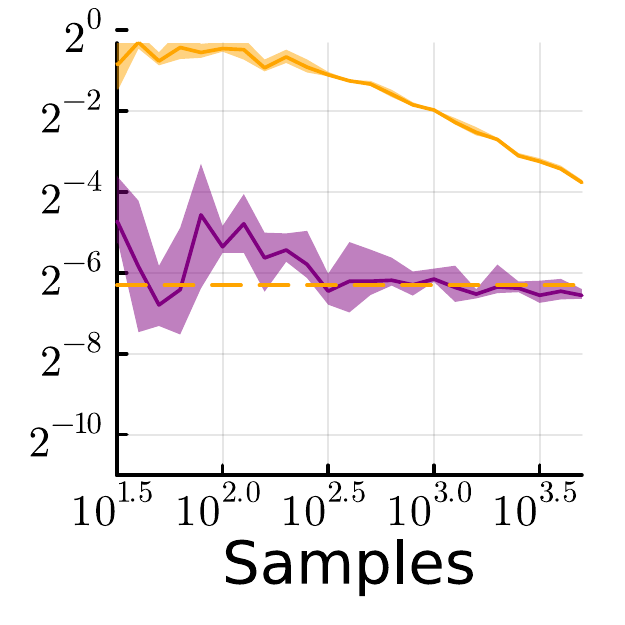}
        \put (-8,90) {\huge \textbf{b}}
    \end{overpic}
    \includegraphics[width=0.23\textwidth]{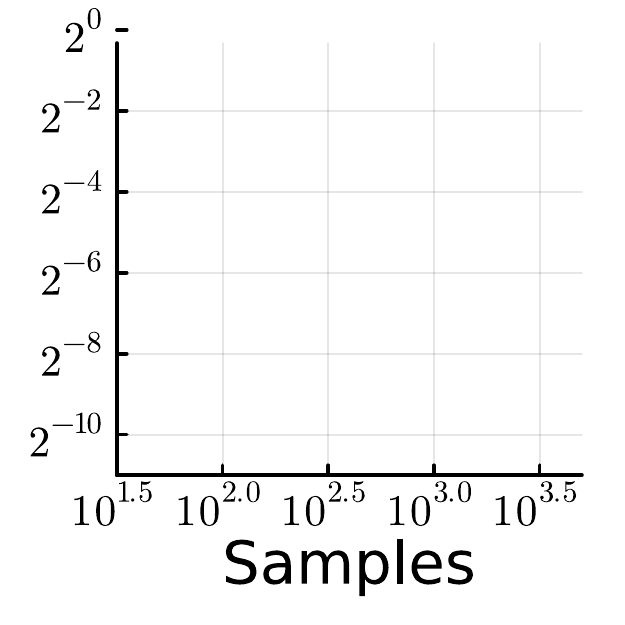}
    \caption{\textbf{Results for calculating transfer entropy from the XY model}. In \textbf{a} the relationship between number of samples used, and transfer entropy estimate is shown. Figures show for past lengths of one (left) and two (right). Here, distance is fixed at one. The dotted histogram line is after the histogram estimate has converged, here using $5 \times 10^6$ samples. \textbf{b} makes a similar comparison, but varying distances of one (left) and two (right), with fixed past length. Again, the dotted line represents the converged histogram method, using a far greater, $5 \times 10^6$, number of samples.}
    \label{fig:te-estimate}
\end{figure}

A similar story is shown in the case of transfer entropy; this is illustrated in Fig.~\ref{fig:te-estimate}. The
histogramming method becomes unworkable for larger values of $\ell$, the length of the past considered and so only a limited set of examples are illustrated. It can again be seen that the method here converges to the value the histogram method eventually approaches, but for far less data.

\section{Discussion}

This paper describes and demonstrates a model-free estimator for conditional mutual information. As a metric-based approach it performs well for high-dimensional data in a situation where other model-free approaches require huge data sets. The estimator is very straight-forward, it relies on finding sets of nearest points, in practice by sorting rows of a distance matrix, and counting the sizes of the intersections of these sets.

This approach relies on a metric. It needs a \q{meaningful} distance measure on each of the outcome spaces for the random variable. If the metric is unsuitable information may be lost: the estimator only estimates the information in the distance matrix. In some applications appropriate distance measures are well understood, this approach, for example, came from neuroscience. In neuroscience there is an interest in the information content of spike trains, the sequences of voltage pulses that propagate information from neuron to neuron. The space of spike trains has good metrics but no coordinates \cite{Victor1997,Rossum2001}. For other data an appropriate metric may be less obvious. Nonetheless, for data with coordinates there are always induced metrics, such as the obvious $L^2$-metric and this proximity structure is often implicitly exploited in estimation approaches, such as using a histogram. In the estimator here, there is a broader choice of metric, since the metric is not linked to the coordinates.

It has been assumed throughout that the smoothing parameter $h$ is the same for each random variable so that each of the balls $B_\x$, $B_\y$ and $B_\z$ has the same volume. In fact, this is not required, separate values $h_\x$, $h_\y$ and $h_\z$ could be used. Maximising over all three could only increase the estimated conditional mutual information; in cases where the the information is very low, this could lead to an over estimate but this freedom to have different smoothing parameters may be needed to get a good estimate in examples where the three distributions are very different.

Another important factor in the application of this method is the computational expense. The estimator appears to work effectively on a comparably low amount of data but it does require the calculation of the distance matrices for the observations in $\mathcal{X}$, $\mathcal{Y}$ and $\mathcal{Z}$ and, significantly, it involves sorting each row of these matrices.  This quickly becomes the most time consuming step as the amount of data increases so the computation time increases as $n\log{n}$.

\subsection*{Applications}

As well as transfer entropy, another measure based on conditional mutual information is interaction information \cite{McGill1954}. The interaction information generalizes the mutual information to more than two variables, it has been applied to problems in biology \cite{MooreEtAl2006,MargolinEtAl2010,LevineWeinstein2014} and has interesting mathematical properties with potential application to the study of intelligence \cite{Bell2003}. For three random variables it is
\begin{equation}
  I(X,Y,Z)=I(X,Y)-I(X,Y|Z)
\end{equation}
with the definition for higher numbers of variables given recursively:
\begin{align}
  I( X_1,X_2,\ldots,X_n)=&
  I(X_1,X_2,\ldots,X_{n-1})\\ & -I(X_1,X_2,\ldots,X_{n-1}|X_n) \nonumber
\end{align}
Since the formula for $I(X,Y,Z)$ includes the conditional mutual information, the estimator describe here gives an estimator for interaction information:
\begin{equation}
  I(X,Y,Z)\approx \frac{1}{n}\sum_{i}\log{\frac{nh_{\x\y}(i)h_{\y\z}(i)h_{\z\x}(i)}{h_{\x\y\z}(i)}}
\end{equation}
with similar, but increasingly complicated, formulae for interaction information with more and more variables. 

It is hoped that by reducing the amount of data required to calculate conditional mutual information, and by implication, the transfer entropy, this new estimator will unlock other applications of these quantities, including applications in data science and machine learning.

\newpage

\appendix

\section{Dealing with draws when counting points}

One issue that does arise is how to deal with draws: what happens if there are multiple points equally far from the seed point and they cannot all be included without the volume exceeding $h/n$? It might seem unlikely that there will be draws given that the distance is generally a real number, however, that does not account for the idiosyncrasy of data. In the example of spike trains in neuroscience, some of the data points might correspond to empty spike trains and these will all be the same distance from each other. As another example, again related to neuroscience, one crude metric gives the difference in the number of spikes as the distance between two spike trains, this discrete metric will obviously lend itself to draws.

Draws can be dealt with by counting points fractionally. Consider drawing a ball around $x_i$, $y_i$ or $z_i$ aiming to include $h$ points. Let $b$ be the number of points on the boundary: the draws or points that are equidistant. Let $c$ be the points in the interior of the ball. The issue arises when $b+c>h$. This is solved by including all of the interior $c$ points, as well as all of the boundary $b$ points. The boundary points however are weighted at $(h-c)/b$, while interior points, as in the regular case, are weighted one. These weights are taken in to account when counting the sizes of intersections.

\section{Variance of estimates}

% Here, the variances of the estimated conditional mutual information for the Markov tree example are shown, corresponding to Figures \ref{fig:markov-tree-results-1d}, \ref{fig:markov-tree-results-2d}.
\begin{widetext}
    \begin{center}
    \includegraphics[width=0.8\textwidth]{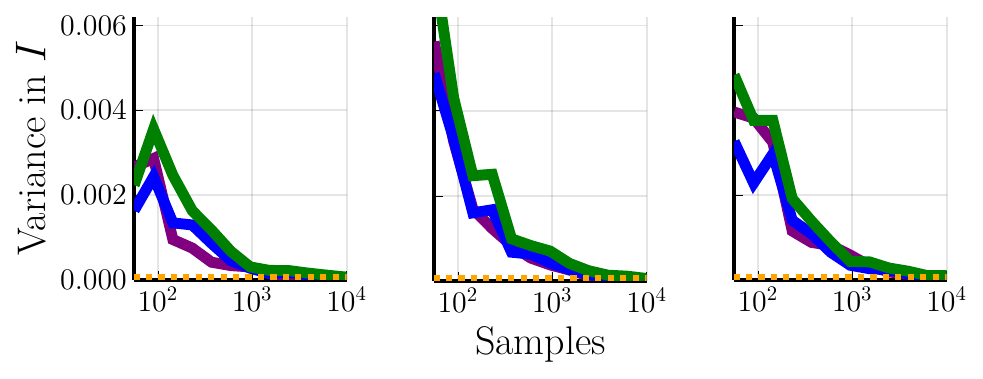} \\
    \includegraphics[width=0.8\textwidth]{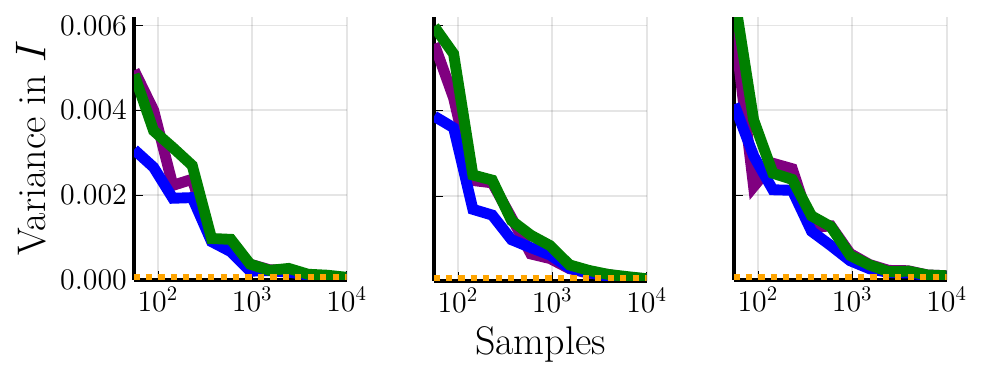}
    \end{center}
Variance of the 50 conditional mutual information estimates in Figures \ref{fig:markov-tree-results-1d}, \ref{fig:markov-tree-results-2d}.
\end{widetext}

% \nocite{*}

\bibliography{bibliography}% Produces the bibliography via BibTeX.

\end{document}